\documentclass[]{aa}
\usepackage[varg]{txfonts}
\usepackage[english]{babel}
\usepackage[flushleft]{threeparttable}
\usepackage{epsfig}
\usepackage{graphicx}
\usepackage{subfig}
\usepackage{natbib} 
\usepackage{siunitx}
\usepackage{amsmath}
\usepackage{geometry}
\usepackage[]{xcolor}

\title{A new view of the corona of classical T Tauri stars:\\
Effects of flaring activity in circumstellar disks}

\author{S. Colombo\inst{\ref{unipa}, \ref{LERMA}, \ref{inaf}} 
	\and S. Orlando\inst{\ref{inaf}}
	\and G. Peres\inst{\ref{unipa},\ref{inaf}}
	\and F. Reale\inst{\ref{unipa},\ref{inaf}}
	\and C. Argiroffi\inst{\ref{unipa},\ref{inaf}}
	\and R. Bonito\inst{\ref{inaf}}
	\and L. Ibgui\inst{\ref{LERMA}}
	\and C. Stehlé\inst{\ref{LERMA}}
	}

\institute{ Dipartimento
	di Fisica \& Chimica Specola Universitaria, Università degli Studi di Palermo, Piazza del
	Parlamento 1, 90143 Palermo, Italy \label{unipa} 
	\and LERMA, Osbervatoire de Paris, Sorbonne Universit\'e, Universit\'e PSL, CNRS, Paris, France \label{LERMA}
    \and INAF-Osservatorio Astronomico di Palermo "G.S. Vaiana", Piazza del Parlamento 1, 90134 Palermo, Italy \label{inaf} }

\begin{document}

 \abstract{Classical T Tauri stars (CTTSs) are young low-mass stellar objects accreting mass from their circumstellar disks. They are characterized by high levels of coronal activity as revealed by X-ray observations. This activity may affect the disk stability and the circumstellar environment.}
{Here we investigate if an intense coronal activity due to flares occurring close to the accretion disk may perturb the inner disk stability, disrupt the inner part of the disk and, possibly, trigger accretion phenomena with rates comparable with those observed.}
{We model a magnetized protostar surrounded by an accretion disk through 3D magnetohydrodinamic simulations. The model takes into account the gravity from the central star, the effects of viscosity in the disk, the thermal conduction (including the effects of heat flux saturation), the radiative losses from optically thin plasma, and a parameterized heating function to trigger the flares. We explore cases characterized by a dipole plus an octupole stellar magnetic field configuration and different density of the disk or by different levels of flaring activity.}
{As a result of the simulated intense flaring activity, we observe the formation of several loops that link the star to the disk; all these loops build up a hot extended corona with an X-ray luminosity comparable with typical values observed in CTTSs. The intense flaring activity close to the disk can strongly perturb the disk stability. The flares trigger overpressure waves which travel through the disk and modify its configuration. Accretion funnels may be triggered by the flaring activity, thus contributing to the mass accretion rate of the star. Accretion rates synthesized from the simulations are in a range between $10^{-10}$ and $10^{-9}M_{\odot}$yr$^{-1}$. The accretion columns can be perturbed by the flares and they can interact with each other, possibly merging together in larger streams. As a result, the accretion pattern can be rather complex: the streams are highly inhomogeneous, with a complex density structure, and clumped.}
 {}

\titlerunning{Coronae of CTTSs: Effects of flaring activity in circumstellar disks}
\authorrunning{S. Colombo et~al.}

\maketitle

\section{Introduction}
Classical T Tauri stars (CTTSs) are young low-mass stars surrounded by a thick quasi-Keplerian disk. According to the largely accepted magnetospheric accretion scenario, the gas of the disk accretes onto the star through accretion funnels \citep{1991Apj...370L..39K}. The disk is truncated internally at a few stellar radii (at the so called truncation radius), where the gas pressure equals the magnetic pressure.  Closer to the star, the magnetic field is strong enough to force the material to move along the magnetic field lines, and to accrete onto the stellar surface. This scenario is well supported by optical and infrared observations \citep{1988ApJ...330..350B,2007prpl.conf..479B}. The accretion process plays an important role during the early phase of stellar formation by regulating the exchange of mass and angular momentum between the star and the disk. Nevertheless, despite the important role of the accretion to understand the physics of stellar formation, there are some points not fully understood. 

CTTSs are also known to be strong X-ray emitters, characterized by an high level of coronal activity. One of the fundamental and most debated issues in this field is the apparent correlation between coronal activity and accretion \citep{2003A&A...397..611F,2005ApJS..160..401P}. In particular CTTSs, presenting active accretion, show an X-ray luminosity that is systematically lower than that observed in Weak line T Tauri Stars (WTTSs), that do not present accreting signatures \citep{1995A&A...297..391N,2009ApJ...699L..35D,2005ApJS..160..401P,2007MNRAS.379L..35G}. 
This evidence suggests that the coronal activity may be influenced by accretion or viceversa. Several scenarios were proposed in the last 15 years to explain if and how the coronal activity is linked to accretion. Some authors suggested that the accretion modulates the X-ray emission through the suppression, disruption or absorption of the coronal magnetic activity \citep{2003A&A...402..277F,2004AJ....127.3537S,2005ApJS..160..401P,2006MNRAS.367..917J,2007MNRAS.379L..35G}.
Other authors suggested that the coronal activity modulates the accretion flow driving the X-ray photoevaporation of the disk material \cite[e.g.][]{2009ApJ...699L..35D}.
\cite{2010ApJ...710.1835B} have suggested that the accretion may even enhance the coronal activity around the region of impact of accretion streams. They proposed a scenario in which the accretion phenomena produces hot plasma which populates the stellar corona, that combined with different magnetic field configurations can be constrained into loops or stellar winds.

Observations show that the X-ray luminosity of a CTTS is, in general, 3-4 orders of magnitude higher than the peak X-ray luminosity of the Sun at present time \citep{2005ApJS..160..353G, 2007A&A...468..379A}. Part of this X-ray emission comes from the heated plasma in the outer part of the stellar corona with temperature from 1 to 100 MK. The plasma heating is presumably due to the strong magnetic activity \citep{1999ARA&A..37..363F} in the form of high energetic flares that are generated from a quick release of energy in proximity of the stellar surface.

In the last years, X-ray observations proved that flares in CTTSs are more energetic and frequent than the solar analogues \citep{2005ApJS..160..469F, 2005ApJS..160..353G, 2007A&A...468..379A, 2010ApJ...717...93A}. The \textit{Chandra} X-ray observatory, during the COUP (\textit{Chandra Orion Ultradeep Project}) observing campaign, revealed numerous flares in the Orion star forming region \citep{2005ApJS..160..469F} with some of them reaching temperatures in excess of $\approx 100$ MK. The analysis of these flares revealed that they are long-lasting and, apparently, they are confined in very long (several stellar radii) magnetic structures that may connect the disk surface to the stellar photosphere (e.g. \citealt{2005ApJS..160..469F, 2016A&A...590A...7L}).  More recently, \cite{2018ApJ...856...51R}, using accurate hydrodynamic simulations of flares confined in magnetic flux tubes, proved unambiguously that some of the flares observed during the COUP campaign are confined in loops extending several stellar radii, possibly connecting the protostar to the disk. 
These flares can be generated by the reconnection of magnetic field lines in the magnetic corona above the disk, as it is often observed in local and global magneto-rotational instability (MRI) simulations of accretion disks \citep[e.g.][]{2000ApJ...534..398M,2011MNRAS.416..416R}; or, by the inflation of the field lines connecting a star with the inner disk, as suggested by other authors \citep[e.g.][]{1997ApJ...489..199G,1999ApJ...524..159G, 2008A&A...478..155B}.

The possible effects of strong flaring activity on the stability of circumstellar disks around CTTSs have been investigated, for the first time, by \cite{2011MNRAS.415.3380O} (hereafter Paper I). These authors developed a 3D magnetohydrodynamics (MHD) model describing the evolution of a flare occurring near the disk around a rotating magnetized star. They explored the case of a single bright flare with energy of the same order of magnitude of those involved in the brightest X-ray flares observed in COUP (\citealt{2005ApJS..160..469F}). They found that the flare produces a hot magnetic loop which links the star to the disk. Moreover the disk is strongly perturbed by the flare: a fraction of the disk material evaporates in proximity of the flaring loop and an overpressure wave originating from the flare, propagates through the disk. When the overpressure reaches the opposite side of the disk, disk material is pushed out to form an intense funnel stream channelled by the magnetic field and accreting mass onto the centralprotostar.

Starting from the results of Paper I, here we study the effects of a storm of flares with small-to-medium intensity (compared to Paper I) occurring in proximity of a disk surrounding a central protostar. The aim is to investigate if the common coronal activity of a young star made by several flares with small-to-medium intensity is able to perturb the disk and to trigger accretion as the single bright flare studied in Paper I. To this end, we adopted the 3D MHD model of the star-disk system presented in Paper I; the model includes the most important physical processes, namely the gravitational effects, the disk viscosity, the radiative losses from optically thin plasma, the magnetic field oriented thermal conduction, and the coronal heating (including heat pulses triggering the flares). We explored cases characterized by different densities of the disk or by different levels of coronal activity.

The paper is structured as follow: in Sect. 2 we describe the physical model and the numerical setup; in Sect. 3 we present the results of the modeling and the comparison with observations; in Sect. 4 we summarize the results and draw our conclusions.

\section{MHD modeling}\label{model}
\begin{figure}
        \centering
        \includegraphics[width=8cm]{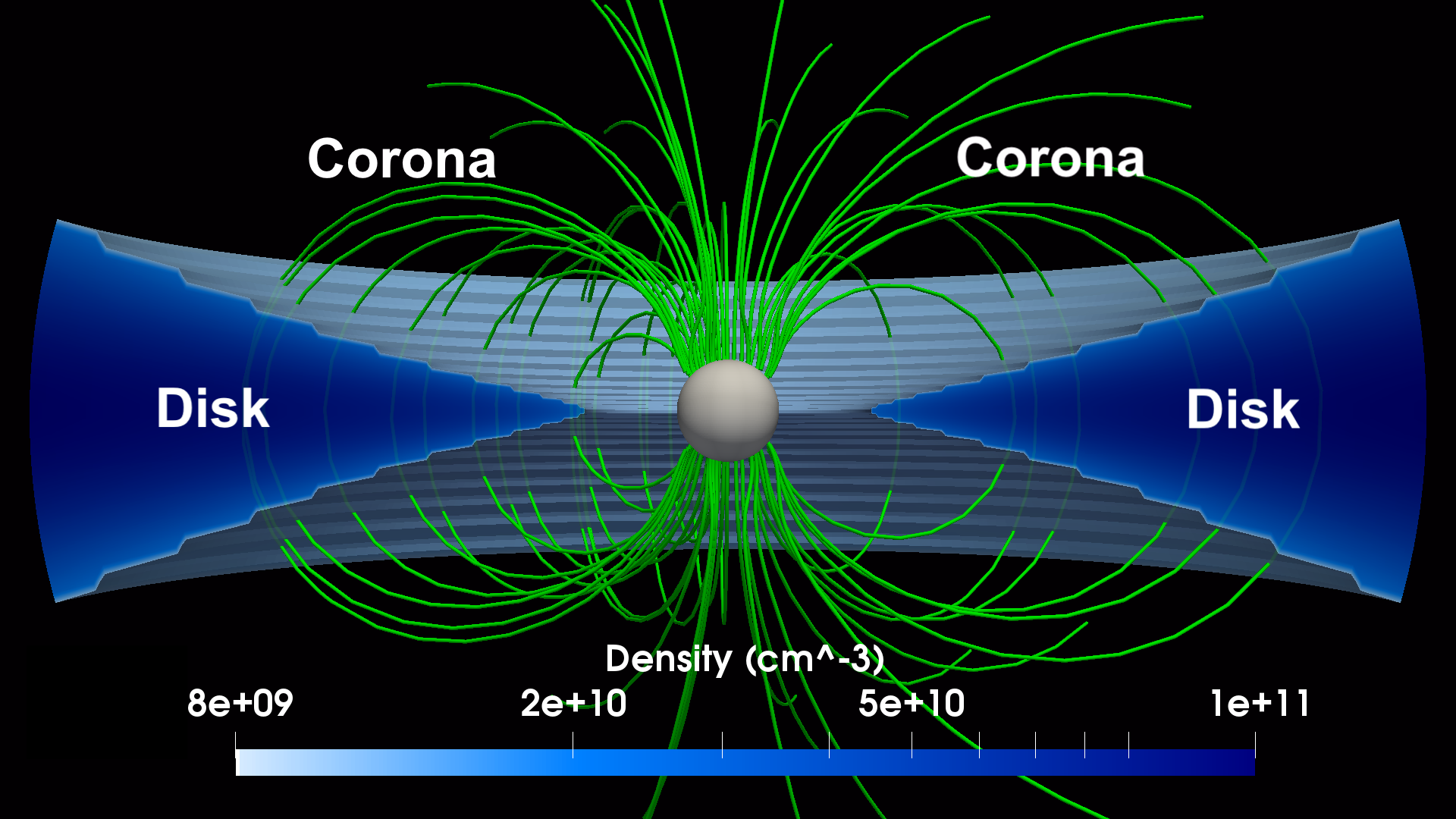}
        \caption{Initial conditions for the reference case. In blue,  in logarithmic scale, the density map of the disk; the green lines are sampled magnetic field lines; the white sphere is the stellar surface that marks a boundary.\label{CI}}
\end{figure}

We adopted the model presented in Paper I, describing a rotating magnetized CTTS surrounded by a thick quasi-Keplerian disk (see Fig.~\ref{CI}), but modified to describe the effects of a storm of small-to-medium flares occurring close to the disk. The CTTS is assumed to have mass $M_\star = 0.8M_\sun$ and radius $R_\star = 2R_\sun$ (see Paper I). The flares occur in proximity of the inner portion of the disk. Initially the magnetosphere is assumed to be force-free, with topology given by a dipole and an octupole \citep{2007MNRAS.380.1297D, 2017A&A...607A..14A}, with both magnetic moments aligned with the rotation axis of the star\footnote{In Paper I, we assumed the magnetic field to be aligned dipole-like.}.
We assumed that the octupole-to-dipole strength ratio is 4, as suggested for the CTTS TW Hya\footnote{This ratio can be different in different young accreting stars (e.g. \citealt{2011AN....332.1027G}).} \citep{2011AN....332.1027G}. The magnetic moments are chosen in order to have a magnetic field strength of the order of $\approx$ 3 kG at the surface of the star \citep{2007MNRAS.380.1297D}. The field ratio determines the topology of the magnetic field lines and, as a consequence, it is expected to influence the dynamical evolution of the accretion streams. The dipole component has a magnetic field strength of about 700 G at the stellar surface and 27 G at the distance of 3 $R_\star$ (in proximity of the disk). At distances larger than $1.7 R_*$, the dipole component dominates and disrupts the inner portion of the disk at a distance of few stellar radii. Closer to the star (at distances below $1.7 R_*$), the octupole component becomes dominant and the accreting plasma is expected to be guided by the magnetic field at higher latitudes than in the pure dipole case (e.g. \citealt{2011MNRAS.411..915R}).

\subsection{Equations}\label{model_eq}
The system is described by solving the time dependent MHD equations in a 3D spherical coordinates system ($R$, $\theta$, $\phi$), including the effects of gravitational force from the star, the viscosity of the disk, the thermal conduction including also the effects of heat flux saturation, the coronal heating (using a phenomenological term composed of a steady state component and a transient component; see Sect. \ref{heating}) and the radiative losses from optically thin plasma. The time-dependent MHD equations are:
\begin{gather}
	\frac{\partial}{\partial t}\rho + \nabla \cdot  \vec{m} =0\\	
	\frac{\partial}{\partial t}\vec{m}+\nabla \cdot (\vec{m}\vec{u} - \vec{B}\vec{B}+ \vec{I}p_t - \vec{\tau }) = \rho\vec{g}\\		
	\begin{split}\frac{\partial}{\partial t}\rho E+\nabla\cdot[(\rho E+p_t)\vec{u}-\vec{B}(\vec{u}\cdot\vec{B})-\vec{u}\cdot \vec{\tau }] = \\& \hspace{-3,5cm}  \vec{m}\cdot\vec{g} - \nabla \cdot F_c - n_e n_H {\Lambda}(T) + Q(R, \theta , \phi , t) \end{split} \label{RL} \\
	\frac{\partial}{\partial t}\vec{B}+\nabla\cdot(\vec{u}\vec{B}-\vec{B}\vec{u})=0
\label{eq_Q}
\end{gather}

\noindent
where: 
\begin{equation}
p_t= P+\frac{\vec{B} \cdot \vec{B}}{2} , \qquad E = \epsilon + \frac {\vec{u} \cdot \vec{u}}{2}+\frac{\vec{B} \cdot \vec{B}}{2\rho}
\end{equation}

\noindent
are the total pressure (thermal and magnetic) and the total gas energy per unit mass (thermal, bulk kinetic and magnetic), $\rho $ is the density, $\vec{m}=\rho \vec{u}$ is the momentum density, $\vec{u}$ is the fluid bulk velocity, $\vec{B}$ is the magnetic field, $\vec{I}$ the identity matrix, $\vec{\tau}$ is the viscous stress tensor treated below in more details, $\vec{g}=-\nabla \Phi _g$ is the gravity acceleration vector, $\Phi_g = GM_\star/R$ is the gravitational potential of a central star of mass $M_\star$ at a distance $R$, $G$ is the gravitational constant, $F_c$ is the heat conductive flux, $n_e$ and $n_H$ are the electron and hydrogen number density, $\Lambda(T)$ is the optically thin radiative losses per unit emission measure, $T$ is the fluid temperature, and $Q(R,\theta,\phi,t)$ is a function of space and time describing the phenomenological heating rate (See Section \ref{heating}).
We use the ideal gas law $P = (\gamma -1)\rho \epsilon$, where $\gamma$ is the adiabatic index and $\epsilon$ is the thermal energy density. The radiative losses are defined for $T > 10^4$K, and are derived with the PINTofALE \citep{2000HEAD....5.2705K} spectral code and with the APED v1.3 atomic lines database assuming metal abundances equal to 0.5 of the solar value \citep{1989GeCoA..53..197A}.

We assumed the viscosity to be effective only in the circumstellar disk and negligible in the extended stellar corona. The transition between corona and disk is outlined through a passive tracer ($C_{disk}$) that is passively advected in the same manner as density (see Paper I for more details). The tracer is initialized with $C_{disk}=1$ in the disk region and $C_{disk}=0$ elsewhere. During the system evolution, the disk material and the corona mix together, leading to regions with $0 < C_{disk} < 1$. The viscosity works only in regions with $C_{disk} > 0.99$, i.e. in zones consisting of more than the 99\% of disk material. The viscous tensor is defined as:

\begin{equation}
\tau = \eta_v \left[(\nabla \vec{u} ) + (\nabla \vec{u})^T
-\frac{2}{3}(\nabla \cdot \vec{u})\vec{I} \right]
\end{equation}
\noindent
where $\eta_v =\nu_v \rho$ is the dynamic viscosity, and $\nu_v$ is the kinematic viscosity. The accepted scenario considers, as major responsible for the losses of angular momentum, the turbulence in the disk \citep{1973A&A....24..337S}, possibly triggered by MRI \citep{1991ApJ...376..214B, 1998RvMP...70....1B}. Given the difficulties in the description of the phenomenon and the poor knowledge of its details, the efficiency of angular momentum transport within the disk is, in general, described through a phenomenological viscous term, modulated via the Shakura–Sunyaev $\alpha$-parameter (e.g. \citealt{2002ApJ...578..420R}, Paper I). As in Paper I,
the kinematic viscosity is expressed as $\nu_v = {\alpha c_s^2}/{\Omega_K}$, where $c_s$ is the isothermal sound speed, $\Omega_K$ is the Keplerian angular velocity, and $\alpha < 1$ is a dimensionless parameter regulating the efficiency of angular momentum transport within the disk. Simulations of Keplerian disks indicate that the turbulence-enhanced stress tensor, responsible for the outward transport of energy and angular momentum, has a typical dimensionless value ranging between $10^{-3}$ and $0.6$ \citep{2003ARA&A..41..555B}.
In our simulations, we assumed $\alpha = 0.02$ \citep[according to][]{2002ApJ...578..420R}.

Due to the presence of the stellar magnetic field, the thermal conduction is anisotropic, and highly reduced in the direction transverse to the magnetic field. We splitted the thermal conduction into two components, one along and the other across the magnetic field lines, $F_c = F_{\parallel}i+F_{\perp}j$. To allow for smooth transition between the classical and saturated conduction regimes, we followed \cite{1993ApJ...404..625D} and described the two components of thermal flux as (see also \citealt{2008ApJ...678..274O}):

\begin{gather}
 F_{\parallel} = \left(\frac{1}{\left[q_{spi}\right]_{\parallel}} +\frac{1}{\left[q_{sat}\right]_{\parallel}} \right)^{-1} \\
 F_{\perp} = \left(\frac{1}{\left[q_{spi}\right]_{\perp}} +\frac{1}{\left[q_{sat}\right]_{\perp}} \right)^{-1}
\end{gather}

\noindent
where $\left[q_{spi}\right]_{\parallel}$ and $\left[q_{spi}\right]_{\perp}$ are the classical conductive flux along and across the magnetic field lines, respectively, according to \cite{1962pfig.book.....S}, i.e.,

\begin{gather}
	\left[q_{spi}\right]_{\parallel} = -k_{\parallel}\left[\nabla T\right]_{\parallel} = -9.2\cdot10^{-7} T^{5/2} \left[\nabla T\right]_{\parallel} \\
	\left[q_{spi}\right]_{\perp} = -k_{\perp}\left[\nabla T\right]_{\perp} = -3.3\cdot 10^{-16}n_H^2/(T^{1/2}B^2)\left[\nabla T\right]_{\perp}
\end{gather}

\noindent
where $k_{\parallel}$ and $k_{\perp}$ are both in units of erg\,K$^{-1}$s$^{-1}$cm$^{-1}$ and $\left[\nabla T\right]_{\parallel}$ and $\left[\nabla T\right]_{\perp}$ are the thermal gradients along and across the magnetic field lines.  If the spatial scale of the temperature characteristic change becomes too short as compared to the electron mean free path, the heat flux saturates and the conductive flux along and across the magnetic field lines can be described as \citep{1977ApJ...211..135C}:

\begin{gather}
	\left[q_{sat}\right]_{\parallel} = -\rm{sign}(\left[\nabla T\right]_{\parallel}) 5 \phi \rho c_s^3 \\
	\left[q_{sat}\right]_{\perp} = -\rm{sign}(\left[\nabla T\right]_{\perp}) 5 \phi \rho c_s^3
\end{gather}

\noindent
where $c_s$ is the isothermal sound speed, and $\phi$, called flux limit factor, is a free parameter between 0 and 1 \citep{1984ApJ...277..605G}. For this work, $\phi=1$ as suggested for stellar coronae \citep{1989ApJ...336..979B,2002A&A...392..735F}.

The calculations were performed using PLUTO, a modular, Godunov-type, code for astrophysical plasmas \citep{2007ApJS..170..228M}. The code is designed to use parallel computers using Message Passage Interface (MPI) libraries. The MHD equations are solved using the MHD module available in PLUTO with the Harten-Lax-van Leer Riemann solver. The time evolution is solved using a second order Runge-Kutta method. The evolution of the magnetic field is calculated using the constrained transport method \citep{1999JCoPh.149..270B}, which mantains the solenoidal condition at machine accuracy. We adopted the ``magnetic field-splitting'' technique (\citealt{1994JCoPh.111..381T, 1999JCoPh.154..284P,2009A&A...508.1117Z}), by splitting the total magnetic field into a contribution coming from the background stellar magnetic field and a deviation from this initial field; then, only the latter component is computed numerically. The radiative losses $\Lambda$ are calculated at the temperature of interest using a table lookup and interpolation method. The thermal conduction is treated with a super-time stepping technique, the superstep consists of a certain number of substeps that are properly chosen for optimization and stability, depending on the diffusion coefficient, grid size and free parameter $\nu < 1$ \citep{CNM:CNM950}. The viscosity is solved with an explicit scheme, using a second-order finite difference approximation for the dissipative fluxes. 
\subsection{Initial and boundary conditions} As in Paper I, we adopted the initial conditions introduced by \cite{2002ApJ...578..420R}, describing the star–disk system in quiescent configuration. In particular, the initial corona and disk are set in order to satisfy mechanical equilibrium among centrifugal, gravitational and pressure gradient forces (see Paper I and \citealt{2002ApJ...578..420R} for more details); all the plasma is barotropic and the disk and the corona are both isothermal with temperatures $T_d$ and $T_c$, respectively. The stellar rotation axis is normal to the disk mid-plane, and the protostar is characterized by a rotational period of 9.2\,d (see Paper I). The isothermal disk is at $T_d = 10^4$ K and dense (see Table 1); the disk rotates with Keplerian velocity about a rotational axis aligned with the magnetic dipole. In the initial condition the disk is truncated at the radius $R_d$, where the ram pressure of the disk plasma is equal to the magnetic pressure. 
In our setup the 
corotation radius (where the disk rotates with the same angular velocity of the protostar) is $R_{co} = 8.6 R_\star$ (see Paper I). The corona is initially isothermal with temperature $T_c = 4$ MK (see Paper I).

We simulated one half of the whole spatial domain in spherical geometry. Thus, the computational domain extends between $R_{min} = R_{\star}$ and $R_{max} = 14 R_{\star}$ in the radial coordinates, between $\theta_{min} = \ang{5} $ and $\theta_{max} =\ang{174}$ in the angular coordinate $\theta$ and between $\phi_{min} = 0 $ and $ \phi_{max} = \ang{180}$ in the angular coordinate $\phi$. The inner and outer $\theta$ boundaries do not correspond to the rotational axis to avoid very small $\delta\phi$ that increase the computational cost and add no significant insight (see Paper I for more details). The radial coordinate is discretized using a logarithmic grid with 128 points with the mesh size increasing with $R$, so as to have a high resolution close to the stellar surface $\Delta R_{min} \approx 3\times 10^9$cm and a low resolution in the outer part $\Delta R_{max} \approx 4\times 10^{10} $cm. The $\theta$ and $\phi$ coordinates grids are uniformly sampled and composed of 128 points each corresponding to the resolution of $\Delta \theta \approx \ang{1.3}$ and $\Delta\phi\approx\ang{1.4}$, respectively

Our choice to consider only one half of the whole spatial domain is aimed at reducing the computational cost due to the complex dynamics of flares which requires to include the radiative cooling and the thermal conduction. The main consequence of this choice is to introduce a long-wavelength cut-off on the perturbation spectrum of the disk. For instance, long-wavelength modes participating in the development of instability at the disk truncation radius are not represented in our model. Nevertheless, we expect that these particular perturbations are not likely to dominate the system evolution and, also, the study of these perturbations is out of the scope of the present study.

The radial internal boundary condition is defined assuming that the infalling material passes through the surface of the star as in \cite{2002ApJ...578..420R} thus ignoring the dynamic of the plasma after it impacts onto the stellar surface. A zero-gradient boundary condition is set at $R_{max}$ and at the boundaries of $\theta_{min}$ and $\theta_{max}$. Periodic boundary conditions are set at $\phi_{min}$ and $\phi_{max}$.

\subsection{Coronal heating and flaring activity \label{Qterm}}
\label{heating}

The term $Q(R,\theta, \phi, t)$ in Eq. 3 is a phenomenological heating function. It is prescribed as a stationary component, plus a transient component in analogy with the function proposed by \cite{2017A&A...600A.105B}. The former is chosen to balance exactly the radiative losses below 1 MK in the corona,
and maintains an initial quasi-stationary extended tenuous corona.

The transient component describes a storm of small-to-medium flares occurring in proximity of the disk. This component is expected to mimic the flaring activity that may be driven by a stressed field configuration resulting from the twisting of magnetic field lines induced by the differential rotation of the inner rim of the disk and the stellar photosphere \citep{1997Sci...277.1475S}.

Each flare is triggered by injecting a localized release of energy that produces a heat pulse. All the pulses have a 3D Gaussian spatial distribution with a width of $\sigma = 2 \times 10^{10}$cm.  The pulses are randomly distributed in space close to the disk surface at radial distances between the truncation radius and the corotation radius. The flares have a randomly generated timing in order to achieve an average frequency of flare per hour either 1 or 4. Also the total energy released in a single pulse is randomized and, in particular, it ranges between $10^{32}$ and $10^{34}$ erg.  The total time duration of each pulse is \text{300 s}, afterwards the pulse is switched off. The short duration of the pulses was chosen to describe an impulsive release of energy. Tests performed in Paper I proved that a duration of 300 s represents the minimum pulse duration that can be managed by the code. The time evolution of pulse intensity consists of three equally spaced phases of 100 s each: a linearly increasing ramp, a steady part, and a linearly decreasing ramp.

\section{Results\label{results}}
Our model solutions depend on a number of physical parameters, among which the most notable are the density of the disk and the frequency of flares. In the light of this, we considered, as reference case (run FL-REF in Table 1) the disk configuration investigated in Paper I, characterized by a maximum density in the equatorial plane of $2.34\times 10^{10}$~cm$^{-3}$, and a frequency of flares of four per hour. Then we explored the case of a disk five times denser than in the reference case (run FL-HD), and the case of a frequency of flares four times lower than in the reference case (run FL-LF). We, also, considered an additional simulation identical to the run FL-HD but without flares (run NF), to highlight the role played by the flares in the formation of accretion columns in the timescales covered. We considered the case with the densest disk because we expect a more efficient interaction between disk and stellar magnetosphere, that helps the formation of accretion columns. In fact, the simulation FL-HD is the one that generates the highest accretion rates ($\approx 10^{-9}\,M_{\odot}$~year$^{-1}$; see Fig. 8). Table~\ref{tab1} summarizes the various simulations and the main parameters characteristic of each run: $\rho_d$ is the maximum density of the disk in the equatorial plane, $\rho_c$ is the density of the corona close to the disk, and $F_{\rm fl}$ is the frequency of flares.

\qquad
\begin{table}
\centering
\caption{Model parameters defining the initial conditions of the 3D simulations.}
\begin{tabular}{c c c c}
	\hline \hline 
	Run  & $\rho_d$ [cm$^{-3}$] & $\rho_c$ [cm$^{-3}$] & $F_{\rm fl}$ [flares/hour]\\
	\hline  
	FL-REF & $2.34 \times 10^{10}$ & $9.37 \times 10^{7}$ & 4   \\
	FL-HD  & $9.37 \times 10^{10}$ & $9.37 \times 10^{7}$ & 4   \\
	FL-LF  & $2.34 \times 10^{10}$ & $9.37 \times 10^{7}$ & 1 \\
	NF     & $9.37 \times 10^{10}$ & $9.37 \times 10^{7}$ & 0 \\
	\hline
\end{tabular}
\label{tab1}
\end{table}

\begin{figure*}
        \centering
        \includegraphics[width=18cm]{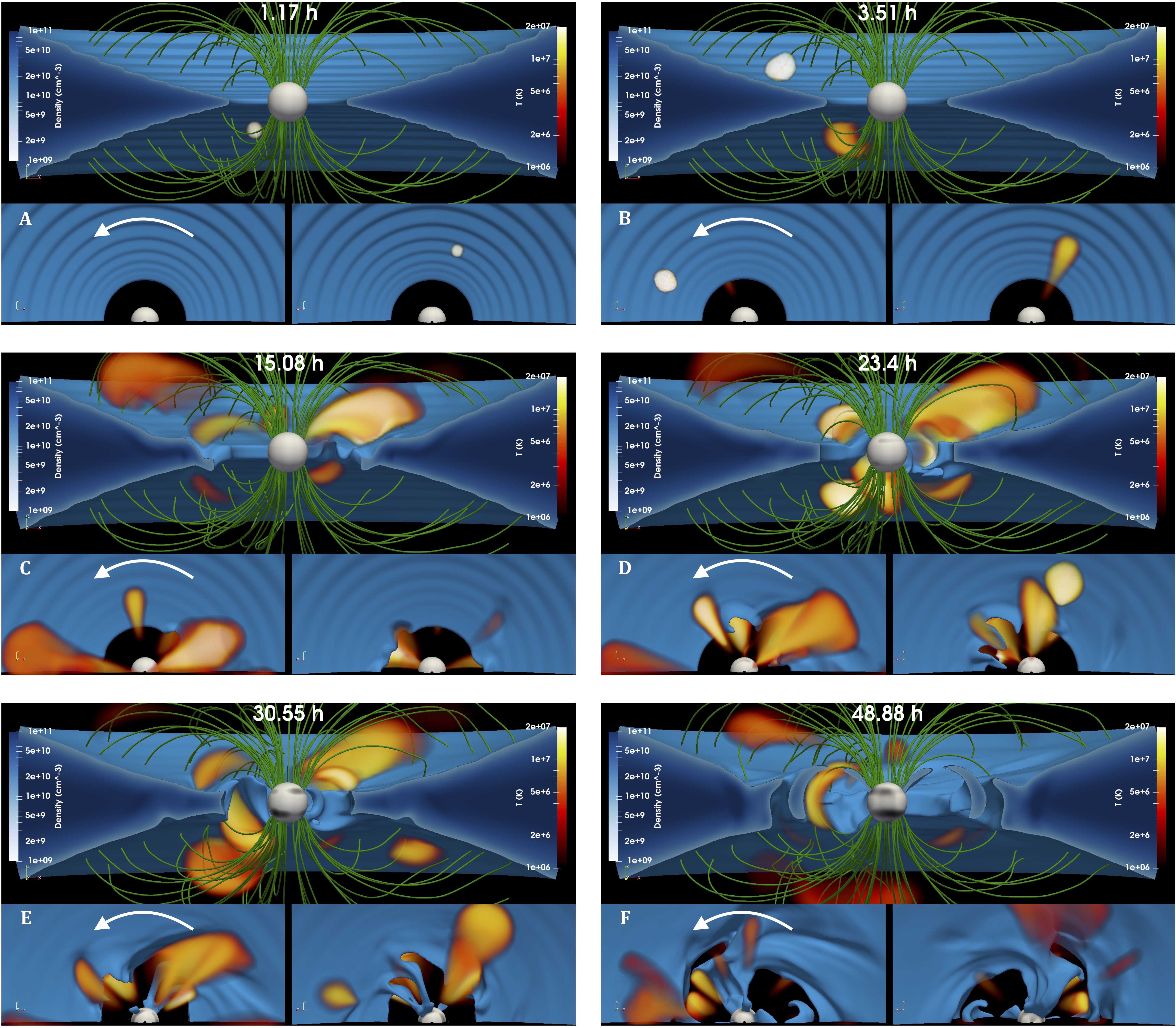}
	\caption{Evolution of the reference case FL-REF. Each of the six panels portrays a snapshot at a given time after the initial condition (defined in Sect. \ref{model} in Fig. \ref{CI}) at 1.17 h (A), 3.51 h (B), 15.08 h (C), 23.4 h (D), 30.55 h (E), 48.88 h (F). Each panel is composed of three images: on top an edge-on view of the 3D system, on bottom left and right the two pole on views. The cutaway views of the star-disk system show the mass density
	(blue) and sampled magnetic field lines (green) at different
	times.  A 3D volume rendering of the plasma temperature is
	over-plotted in log scale in each image and shows the flaring
	loops (red-yellow), linking the inner part of the disk with
	the central protostar. The color-coded density logarithmic scale is shown on the left of each panel, the analogously coded temperature scale is on the right. The white arrows indicate the direction of rotation of the system. The physical time since the start of the evolution is shown at the top center of each panel. \label{RFC}}
\end{figure*}

In this work the main focus is to explore the role played by the flares in perturbing the accretion disk and, possibly, in triggering accretion on timescales shorter than those if flares were not present. For this reason, we stopped the simulations when the accretion rates reach a quasi-stationary regime after the initial sudden and steep rise (see Sect. 3.3); our simulations cover about three days of evolution ($\approx 1/3$ of the rotational period of the inner disk). 

We followed the evolution of the system without flares (i.e. run NF) for about 4 days (i.e. on a timescale larger than that covered by the other simulations). We note that the initial conditions adopted provide quasi-equilibrium. Initially the disk is truncated at $2.86 R_\star$. Once the simulation starts, the magnetic field lines of the slowly-rotating magnetosphere threading the disk exert a torque on the faster rotating disk. As a result, the inner part of the disk moves inward on a timescale which is shorter than the viscous timescale and matter gradually accumulates in proximity of the truncation radius located at $R_d \approx 2.1 R_\star$ at the end of the simulation. As shown by long-term 3D simulations (e.g. \citealt{2002ApJ...578..420R,2011MNRAS.416..416R,2008MNRAS.386..673K}), this process of inward motion of matter leads at some point to accretion onto the star without any need of flares. However, run NF shows that this initial torque helps to bring matter towards the magnetosphere of the star but no accretion stream develops in the timescale considered. By the end of the simulation, there is a hint of mass accretion which leads to rates of the order of $10^{-10}\, M_{\odot}$~year$^{-1}$ (about an order of magnitude lower than in the corresponding simulation with flares; see Fig. 8). We conclude that higher levels of accretion driven by the magnetic torque in our star-disk system start on timescales longer than $\approx 4$~days. In the following, we discuss in detail the simulations including the effect of flares. In the light of the results of run NF, we are confident that disk perturbations and the accretion rates found in the timescale considered are entirely due to the flaring activity.

\subsection{The reference case \label{refsect}}

For the reference case FL-REF, we followed the evolution of the flaring
activity on the star-disk system for approximately 3 days (i.e. when a stationary regime is reached). 
Fig. \ref{RFC} shows representative frames to describe this evolution, a movie showing the entire evolution is provided as on-line material \footnote{The movie shows the evolution of FL-REF. Each frame is composed of three panels: on top an edge-on view of the system, on bottom left and right the two pole on views. The movie shows the mass density (blue) and sampled magnetic field lines (green). A 3D volume rendering of the plasma temperature is over-plotted in log scale and shows the flaring loops (red-yellow). }(Movie 1).

As discussed in Sect.~\ref{heating}, the heat pulses triggering the
flares are injected into the system at randomly chosen locations
above and below the disk, through the phenomenological term $Q(R,
\theta, \phi, t)$ in Eq.(3) We found that the evolution
of each flare is analogous to the evolution of the single bright
flare analyzed in Paper I. For example this is evident for the first flare in
run FL-REF, occurring after $\approx 1$ hour since the initial
condition (see lower right panel in Fig.\ref{RFC} A-B and Movie 1), and not perturbed by other flares
during its entire lifetime.

The initial heat pulse determines a local increase of plasma pressure
and temperature close to the disk surface. As a result, the disk
material is heated up and rapidly expands through the overlying hot
and tenuous corona, resulting in a strong evaporation front. The
disk evaporation is sustained by thermal conduction from the outer
layers of hot plasma even after the heat pulse is over. The fraction
of this hot expanding material closer to the protostar is
confined by the magnetic field due to $\beta < 1$ and forms
a hot (temperature of about $10^7-10^9$~K) magnetic loop of length of the
order of $10^{11}$~cm that links the disk to the stellar surface. The
magnetic-field-oriented thermal conduction promotes the development
of this hot loop through the formation of a fast thermal front
propagating from the disk along the magnetic field lines toward the
star. The remaining part of the evaporating plasma which is not
channeled in the loop is poorly confined by the magnetic field (due
to $\beta > 1$). Thus it moves away from the system
toward regions with higher $\beta$, carrying away mass and angular
momentum from the system (see below). In any case, the plasma (either magnetically
confined or unconfined) starts to cool down as soon as the heat
deposition is over due to the combined action of radiative losses,
thermal conduction, and plasma expansion. After about 10 hours the
hot loop disappears.

\begin{figure}
        \centering
        \includegraphics[width=9cm]{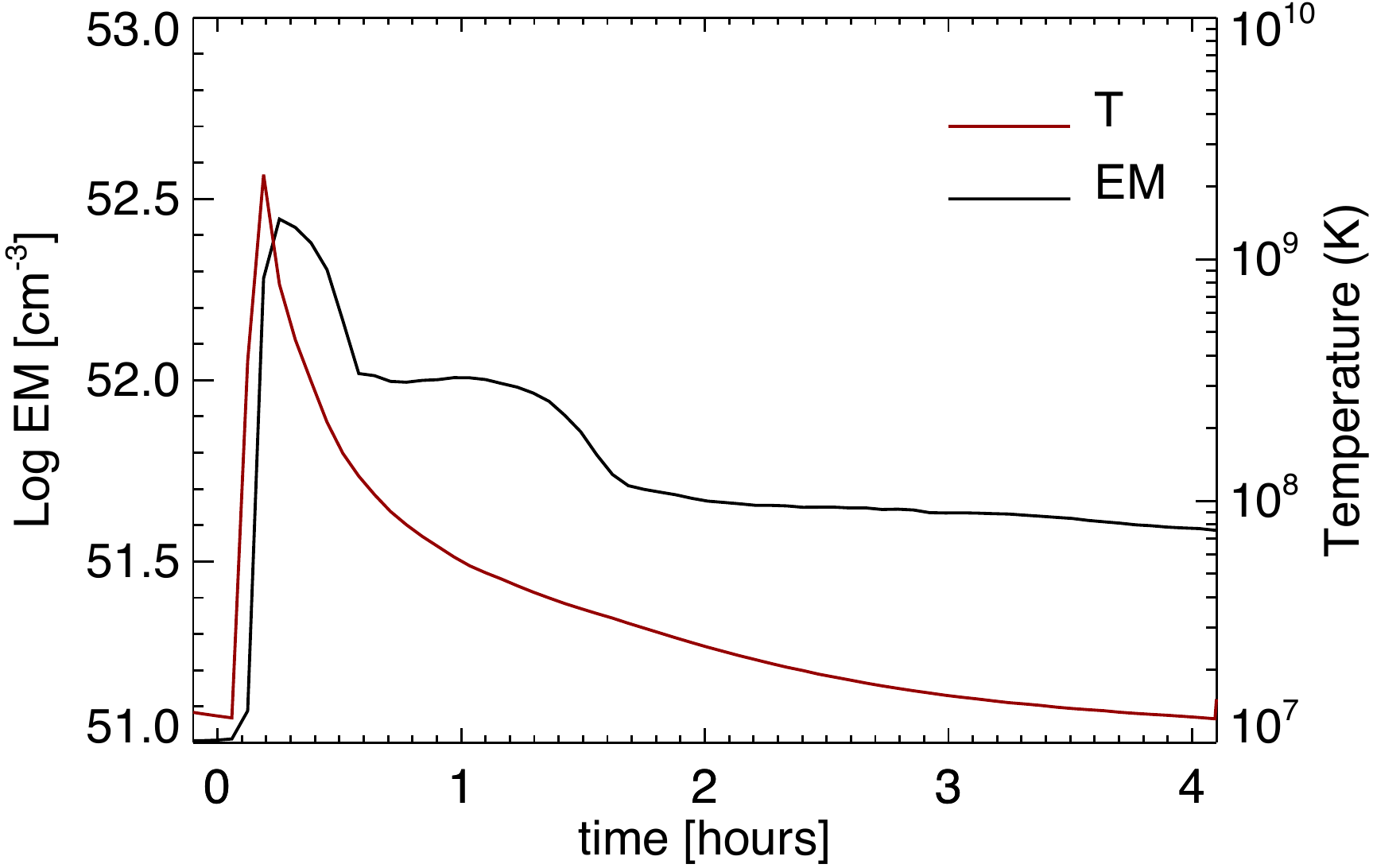}
	\caption{Evolution of the maximum temperature (red line) and
	integrated EM (black line) of the plasma with
	$\log(T) > 6.3$ for the second flare in run FL-REF. The initial time is at the release of the corresponding heat pulse.}
\label{flare}
\end{figure}

Fig.~\ref{flare} shows the evolution of the maximum temperature and
emission measure (EM) of the second flaring loop (see Fig. \ref{RFC} B and Movie 1). This evolution can be considered
to be representative of all the flares simulated here and is analogous to that observed on Sun and stars. We distinguished
two different phases in the evolution: a heating and a cooling
phase. The first starts with the injection of the heat pulse and
lasts for few hundred seconds.  During this phase the temperature
increases very rapidly, reaching a maximum at $\approx 10^9$K. The
disk material evaporates under the effect of the thermal conduction
and the heated plasma expands, filling a magnetic flux tube linking
the disk with the star. As a result, the EM rapidly
increases and peaks at later times, about 6 minutes after
the maximum temperature. At later times, the radiative losses and
the thermal conduction cool down the plasma very efficiently and
the cooling phase starts. Both the temperature and the EM decrease slowly. After about four hours since the heating, the temperature of the flaring plasma is still of the order
of $10^7$~K. The EM develops a second bump about 1 hour after the peak, at odds with the expected evolution on the
base of 1D models of flares (e.g.  \citealt{1988ApJ...328..256R}).
In fact, as discussed in Paper I, each flare of our simulations is
only partially confined by the magnetic field at the loop footpoint
anchored at the disk surface, at variance with 1D models where the
flares are assumed to be fully confined by the magnetic field.
The second bump of EM in Fig.~\ref{flare} originates from the
fast expansion of the unconfined hot plasma in the surrounding
environment. From this point of view, our simulated flares can be
considered to be intermediate between those of models of fully
confined flares (e.g.  \citealt{1988ApJ...328..256R}) and those of
models of unconfined flares (e.g. \citealt{2002A&A...383..952R}).
We also note that the magnetic field in the domain is idealized. In a more realistic case the evolved magnetic field may have a more complex configuration, and may be more intense than in the case studied, due to magnetic field twisting. For this reason we expect to observe a greater confinement in a real case than in the simulations.

In addition to the formation of the hot loop, the initial heat pulse
produces an overpressure wave at the loop footpoint anchored to the
disk which propagates through the disk. In the case of the energetic
flare investigated in Paper I, when the overpressure reaches the
opposite side of the disk, it pushes the material out of the
equilibrium position and drives it into a funnel flow (see Paper I
for more details). Then the gravitational force accelerates the
plasma towards the central star where the accretion stream impacts.
In run FL-REF, however, the flares are less energetic than that in
Paper I. In fact, the overpressure wave generated by the first flare
is not strong enough to perturb significantly the disk and, eventually,
to trigger an accretion event.

After the first, other flares occur close to the disk surface. Each
of them produces a significant amount of hot plasma (with temperatures
up to $\approx 10^9$~K) that is thrown out in the magnetosphere (Fig.
\ref{RFC} C-F and Movie 1). As for the first flare, part of this hot material
is channeled by the magnetic field in flux tubes which form hot
magnetic loops linking the disk to the central protostar. Nevertheless,
most of the evaporated disk material is poorly confined by the
magnetic field (especially in flares occurring at larger distances
from the star) and it escapes from the system, contributing to the
plasma outflow from the star with a mass loss rate
$M_{loss}\approx10^{-10}\,M_{\odot}$~yr$^{-1}$ and angular momentum loss rate $L_{loss} \approx 10^6 - 10^7$ g cm$^2 M_{\odot}^{-1}$~yr$^{-1}$.
As a result, the magnetic field lines are gradually distorted by the escaping plasma. All this hot plasma, either
confined in loops or escaping from the star, forms an extended hot
corona that, once formed, persists until the end of the simulation.

\begin{figure}
        \centering
        \includegraphics[width=9cm]{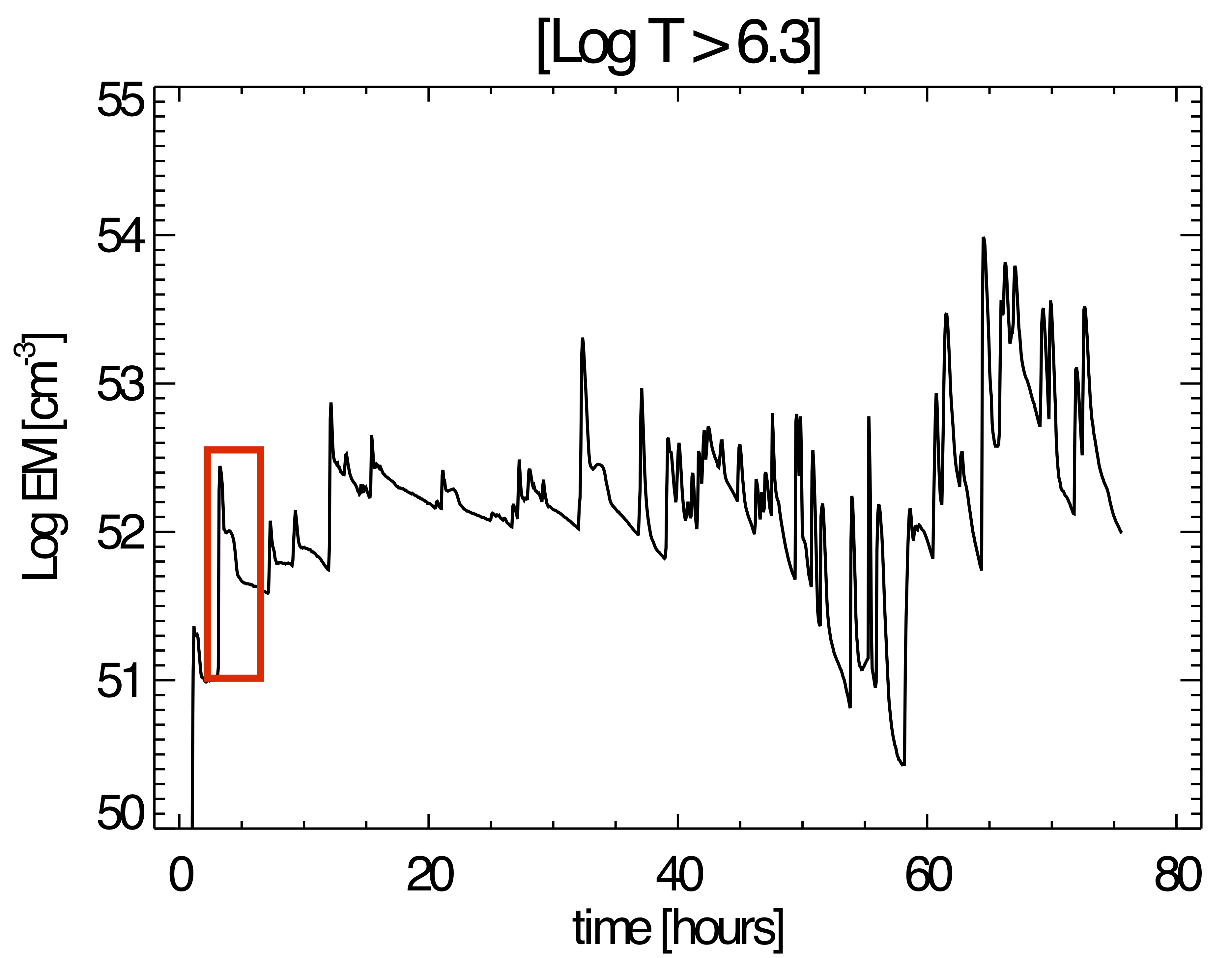}
	\caption{Evolution of the integrated EM for plasma with $\log(T) > 6.3$ for the run FL-REF. The red square is zoomed in Fig.\ref{flare}. }
\label{EMvsT}
\end{figure}

Fig.~\ref{EMvsT} shows the evolution of the integrated EM of the whole system. After the first heat pulse, the emission
measure rapidly increases. In this transient phase, the extended
corona is not developed yet and the effect of each heat pulse is
clearly visible in the curve with the sudden increase and slower
decay of EM.  After the first flares, the amount of plasma
at temperatures greater than 2 MK continues to increase, reaching
a quasi stationary regime after $\approx 10$ hours. At this time
the hot extended corona is well developed. The resulting values of
EM for plasma above 2 MK indicate that the flaring
activity is expected to produce a significant X-ray emission. During
this phase, the pronounced peaks of EM are produced
by the most energetic flares.

We synthesized the X-ray luminosity ($L_X$), in the band [1,10] keV, from the reference case. 
We applied a procedure analogous to the one described in Paper I \citep[see also][]{2017A&A...600A.105B}. 
We found that, during the stationary phase $L_X \approx 10^{30}$erg/s. This value is in the range of luminosities typically inferred from observations \citep{2005ApJS..160..401P}. A detailed study of the evolution and origin of the X-ray emission will be the subject of a future paper.

As for the first flare, each subsequent abrupt release of energy
produces a local increase of plasma pressure in regions of the disk
corresponding to the footpoints of the loops. These local increases
of pressure produce overpressure waves that propagate through the
disk. In general, given the low energy of most of the flares, the
single overpressure wave is not sufficient to perturb significantly
the disk and to trigger an accretion column.  However, the combination
of a series of these overpressures can strongly perturb and distort
the disk, especially in proximity of the truncation radius (see
Fig.~\ref{RFC} F and Movie 1). Eventually, the overpressures can produce mass accretion episodes. In fact after $\approx 15$ hours, several
accretion columns start to develop (Fig.  \ref{RFC} C and Movie 1). The first column impacts onto the stellar surface after $\approx 23$ hours (Fig. \ref{RFC} D and Movie 1). Then, the accretion process continues until the end of the simulation.

The dynamics of the accretion columns is complex, as expected because the corotation radius of our system  ($R_{co}=8.6 R_*$) is much larger than the magnetospheric radius ($\approx 2-3 R_*$). In these conditions, the material accretes in a strongly unstable regime \citep{2016MNRAS.459.2354B}, where one or two unstable filaments (``tongues”) form and rotate approximately with the angular velocity of the inner disk. In our simulation, the system evolution is analogous and  accretion columns develop from the disk surface in regions close to the truncation radius. 
There, the disk material is channelled by the magnetic field in accretion
streams which accelerate towards the central star under the effect
of gravity. Then the accretion streams impact onto the stellar
surface at high latitudes ($> \ang{40}$; see Sect.~\ref{spots}). 
The accretion pattern suggests accretion driven by Rayleigh-Taylor (RT) instability (e.g. \citealt{2016MNRAS.459.2354B}). Indeed, as expected for RT unstable accretion, one can see the formation of small-scale filaments along the edge of the disk, and the formation of several funnels. Their occurrence and shape are typical of unstable tongues: they form frequently and rapidly disappear and initially they are tall and narrow (e.g. \citealt{2008MNRAS.386..673K,2009MNRAS.398..701K,2008ApJ...673L.171R, 2016MNRAS.459.2354B}). In 3D simulations of accretion onto stars with tilted dipole, the instability is triggered and supported by the non-axisymmetry introduced by the dipole and by the pressure gradient force which develops close to the truncation radius due to accumulation of matter by disk viscosity \citep{2012MNRAS.421...63R}.
In our case of a non-tilted dipole-octupole simulated on short timescales, the X-ray flares provide the necessary amount of non-axisymmetry and determine the pressure gradient force to start and to support accretion through instability. In the simulation without flares (run NF), no sign of instability is present in the timescale covered ($\approx 4$~days). Therefore, even if matter would accrete onto the protostar anyway due to MHD processes of the disk-magnetosphere interaction, in our simulations this phenomenon would occur on timescales longer than those covered here. The flares, therefore, provide the conditions to speed up the process on timescales significantly lower (only few hours) than those present in the case without flares.

\begin{figure}
        \centering
	\includegraphics[width=9cm]{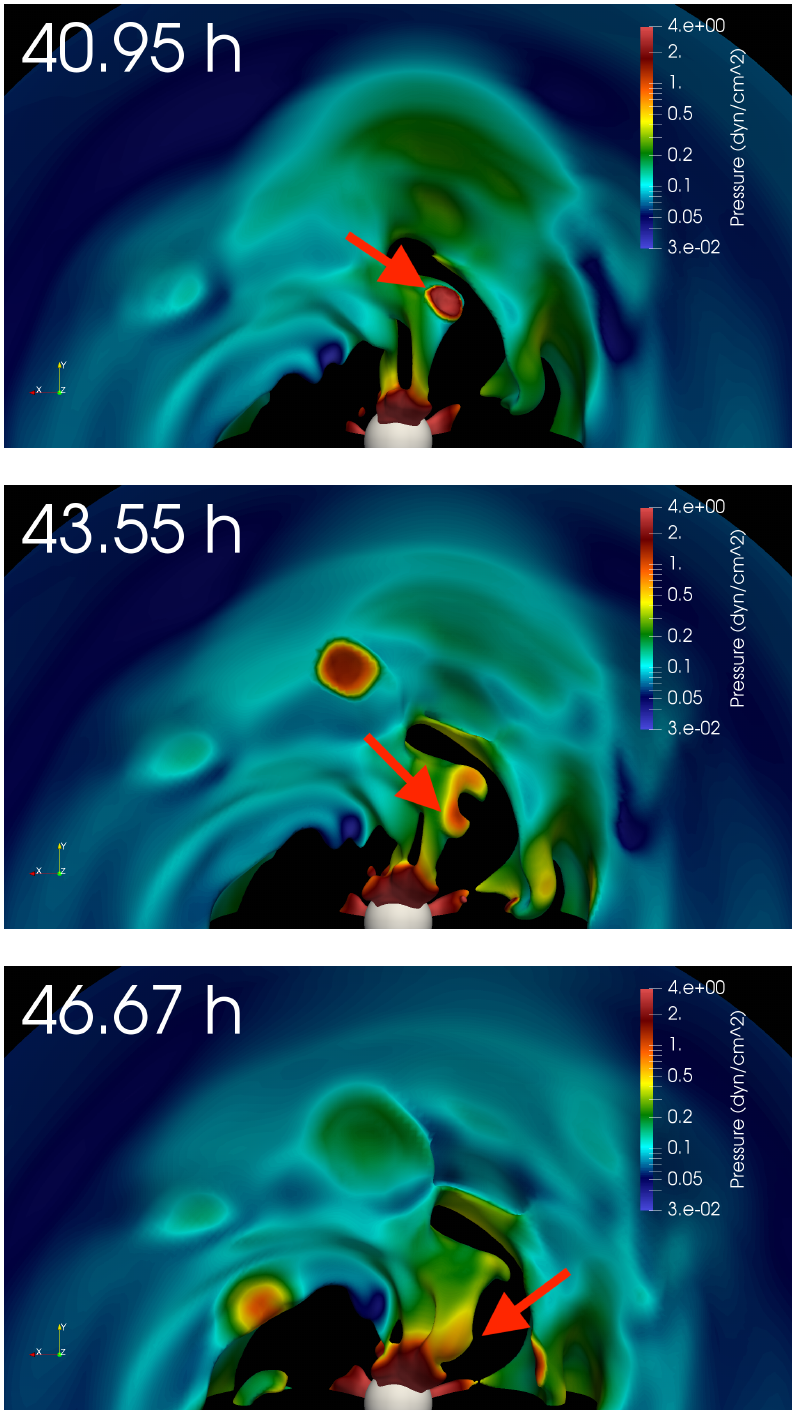}
\caption{Example of the perturbation of an accretion column by nearby flaring activity. The figure shows snapshots displaying isosurfaces where the particle number density equals $5\times~10^{9}$~cm$^{-3}$ in run FL-REF. It is shown the South pole-on view of the star. The isosurfaces, which coincide with the cold and dense disk material, are shown at the labeled times (upper left corner of each panel). The colors give the pressure (in units of dyn cm$^{-2}$) on the isosurface, with the color coding defined on the right of each panel. Upper panel: the red arrow points at flare that perturbs the accretion column (see high pressure region in the disk). Mid and lower panel: the red arrow points the portion of the disk material that perturbs the accretion column. }
\label{perturbation}
\end{figure}

The streams can be heavily perturbed during their lifetime by the flaring activity and by other
streams. For instance, a flare occurring on the opposite side of
the disk may trigger an overpressure wave which may enhance the
accretion rate of the stream. A flare occurring close to the accretion
column may inject more mass in the stream increasing its lifetime
(see Fig.~\ref{perturbation}) or it may produce a perturbation
strong enough to disrupt the base of the stream and, as a consequence,
the whole accretion column. The streams may interact with each
other, merging in larger streams (see Movie 1). 
As a result of this complex dynamic, the streams are highly unstable and time variable; the accretion columns are highly inhomogeneous, structured in density, and clumped (e.g. \citealt{2013A&A...557A..69M, 2016A&A...594A..93C}).  This may support the idea that the persistent low-level hour-timescale variability observed in CTTSs may reflect the internal clumpiness of the streams (\citealt{1996A&A...307..791G, 1998ApJ...494..336S, 2007A&A...463.1017B, 2007A&A...475..891G, 2009ApJ...706..824C}). 

\subsection{Comparison with the other models}
\label{comp_models}

\begin{figure}
        \centering
	\includegraphics[width=8cm]{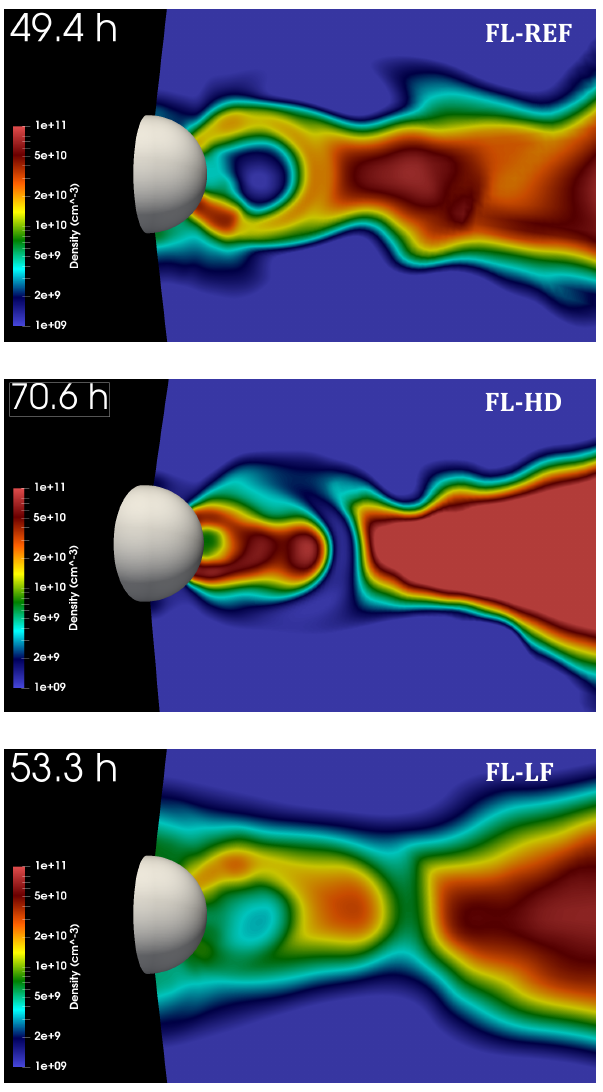}
	\caption{Close up view of the structure of the accretion columns
	for runs FL-REF (top panel), FL-HD (middle panel), and FL-LF
	(bottom panel). Each panel shows a slice in the $(R,z)$
	plane passing through the middle of one of the streams. The white
	hemisphere represents the stellar surface. \label{CD}}
\end{figure}

\begin{figure}
	\centering
 \includegraphics[width=9.5cm]{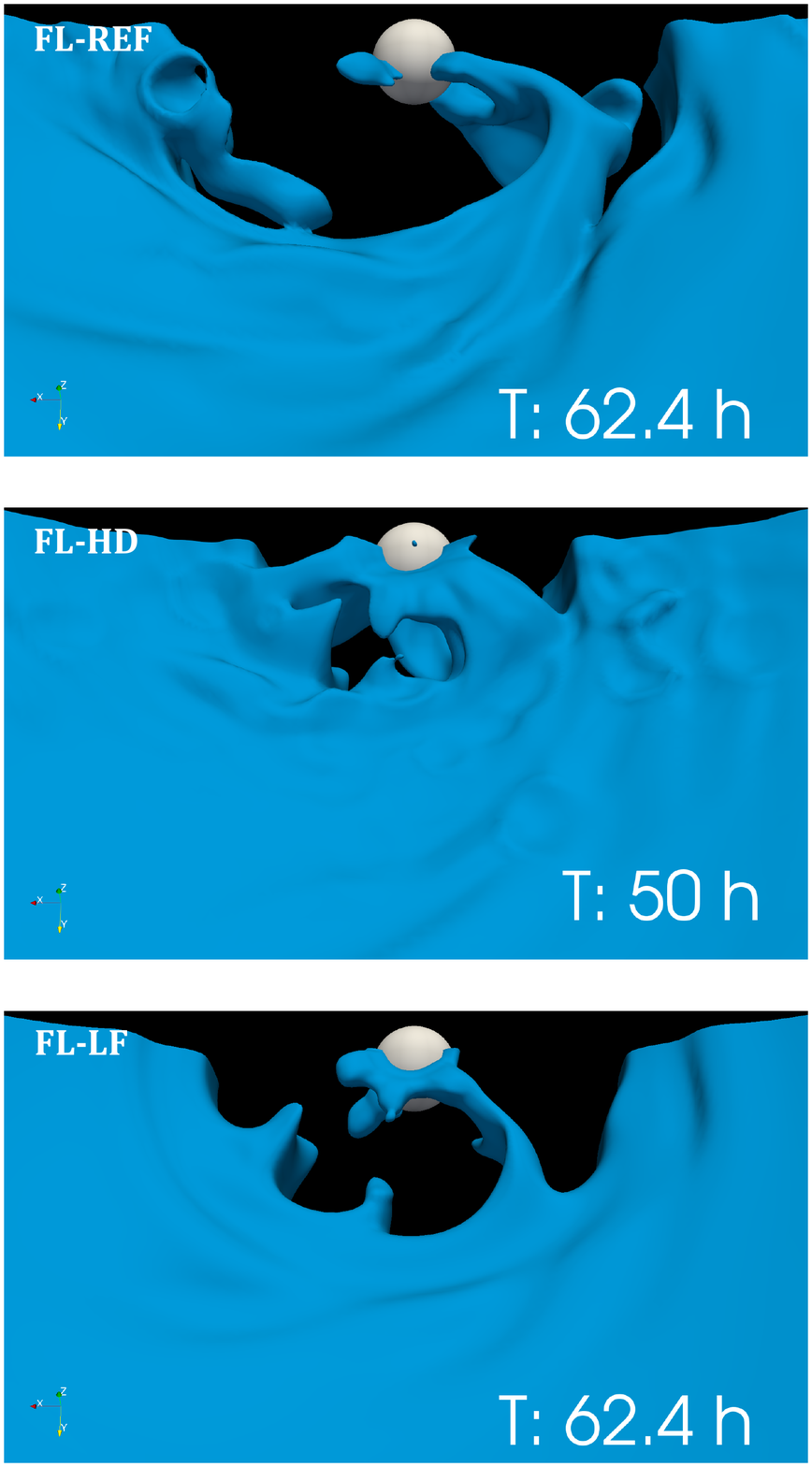}
	\caption{Representative close up views of the inner region of the star-disk
	system for runs FL-REF (top panel), FL-HD (middle panel),
	and FL-LF (bottom panel). In blue the isosurfaces where the
	disk material has particle number density $n=10^{10}$cm$^{-3}$.
	The white sphere in each panel represents the stellar
	surface.}
\label{LF}
\end{figure}

We explored the effects of either the disk density or the flare
frequency on the dynamics of the star-disk system. To this end, we
considered the case of a disk five times denser than in the reference
case (run FL-HD in Table~\ref{tab1}) and with the same sequence of
flares. We found that the evolution in run FL-HD is analogous to
that of the reference case. In particular, the random heat pulses
produce hot plasma in proximity of the disk surface; then part of
this plasma is channeled in hot loops, and the other part escape
from the system. All this hot material forms an extended and
structured hot corona. As for the reference case, the heat pulses
also generate overpressure waves which propagate through the disk,
eventually pushing the disk material out of equilibrium to form
accretion columns. In the case of a denser disk, however, the effects
of these waves moving through the disk are mild in comparison with
run FL-REF. In fact the perturbation of the disk is limited to the
surface of the disk and the latter is less distorted than in FL-REF
(see Figs.~\ref{CD}-\ref{LF}). Nevertheless, the disk material channeled
in accretion streams is, on average, denser than in FL-REF so that
the accretion columns are, in general, denser than in the reference
case (see Fig. \ref{CD}).

As for the dependence of the system evolution on the flare frequency,
we found that the dynamics can be significantly different if the
flare frequency is four times lower than in FL-REF. In this case,
in fact, the system needs more time in order to have a significant
perturbation of the disk that is able to produce mass accretion
onto the central protostar. The dynamics of the accretion columns,
once formed, is similar to that in runs FL-REF and FL-HD. The main
difference is the number of streams, which is lower in run FL-LF
than in the other two cases. This is a direct consequence of the
lower number of flares in run FL-LF, which can contribute to the
disk perturbation with their overpressure waves. After $\approx
60$~hours, only one prominent accretion column is present, whereas in runs FL-REF and FL-HD two or more accretion columns are clearly visible (see Fig.~\ref{LF}). Nevertheless, a significant mass accretion is present in the system at the end of the simulation, even if the flare frequency is lower than in FL-REF.

\subsection{Accretion rates \label{accrsect}}

\begin{figure}
	\centering
       \includegraphics[width=9cm]{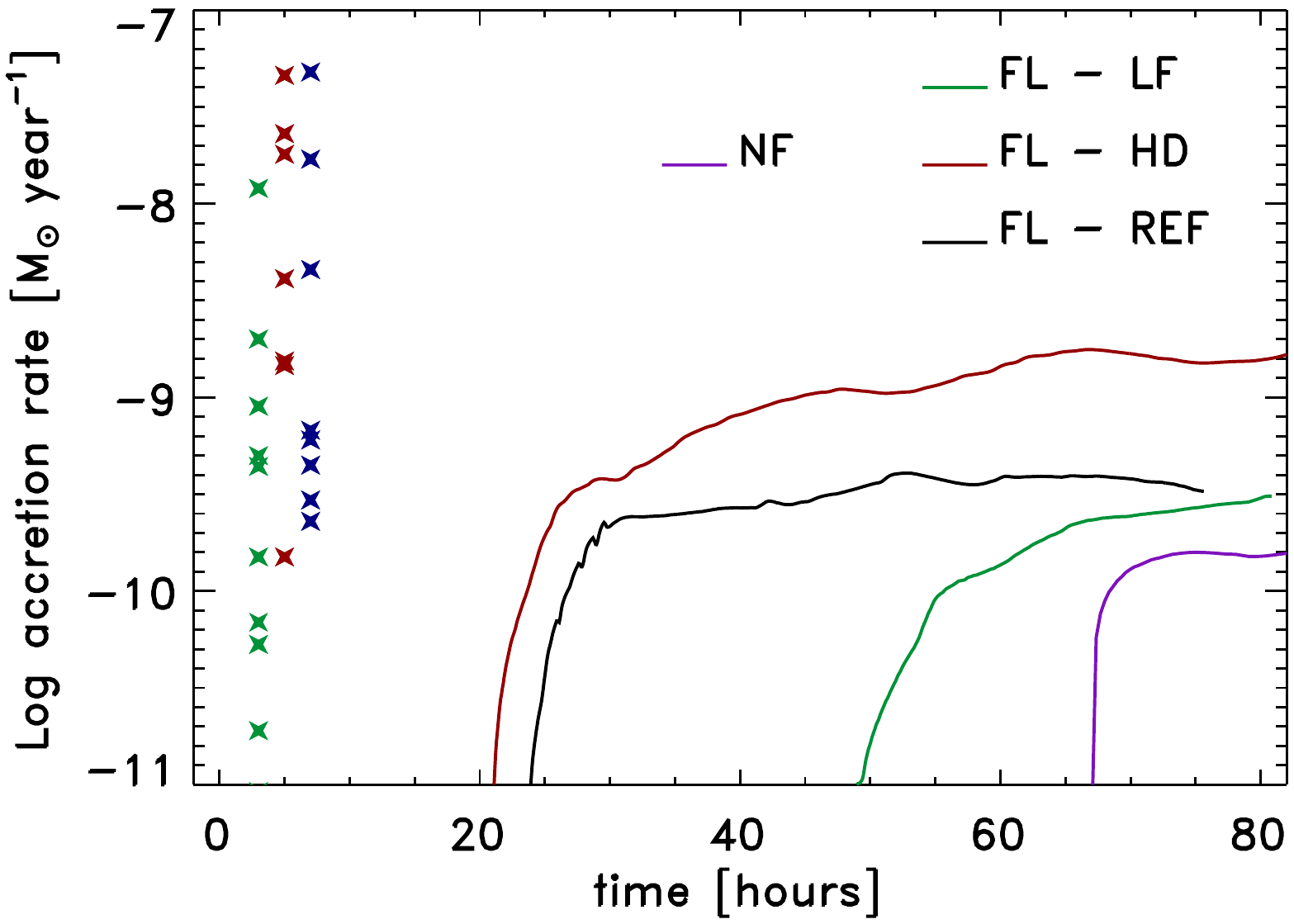}	
	\caption{Evolution of accretion rates synthesized
		 from runs FL-RF (black line), FL-HD (red line),
		 FL-LF (green line), and NF (purple line).
		 The crosses represent the values of mass accretion rates 
		 inferred from optical-near UV observations for a sample of
		 low-mass stars and brown dwarfs \citep[green,][]{2008ApJ...681..594H}, for a sample of solar-mass young accretors \citep[red,][]{2008ApJ...681..594H} and for an
		 X-ray-selected sample of CTTSs
		\citep[blue,][]{2011A&A...526A.104C}; their position on the time axis is choosen for easy comparison with the model results. \label{ACCR_RATE}}
\end{figure}

From the modeling results, we derived the mass accretion rates due
to the various streams triggered by the flares. We considered as
accreting material the amount of mass that goes across the internal
boundary representing the stellar surface, namely the material that
falls onto the central star. Fig. \ref{ACCR_RATE} shows the evolution
of the accretion rates for the four simulations in Table~\ref{tab1}. Qualitatively the three cases with flares show the same trend. After the first stream hits the stellar surface (during the first $20$ hours of evolution; in run FL-LF this happens after about 50 hours), the accretion
rate rapidly increases due to the rapid development of new accretion
columns. This phase is transient and lasts for $\approx 10$ hours
(see Fig.~\ref{RFC} D and Movie 1). After this, the
accretion rate increases more slowly, entering in a phase in which
the number of active accretion columns is roughly constant.

After the transient phase, the three cases show accretion
rates ranging between $\approx 10^{-10}\,M_{\odot}$~yr$^{-1}$ and
$\approx 10^{-9}\,M_{\odot}$~yr$^{-1}$, with the highest values
shown by run FL-HD due to denser streams with respect to the other
two cases, and with the lowest values shown by run FL-LF and FL-REF due to a
less dens disk (see Sect.~\ref{comp_models}). After about
70 hours of evolution, run FL-LF shows accretion rates comparable
to those of run FL-REF. In fact, these two cases differ only for
the frequency of flares and they consider the same density structure
of the disk; thus run FL-LF needs more time to reach the accretion rates
values shown in run FL-REF. We conclude that the accretion rates
depend more on the disk density than on the flare frequency.

The simulation without flares (run NF) does not show accretion during the first $\approx 65~$hours of evolution. In this time the magnetic torque brings matter towards the magnetosphere of the star and, after $\approx 65~$hours, leads to inflow of matter generating a sudden increase in the accretion rate up to $\approx 10^{-10}\,M_{\odot}$~yr$^{-1}$. Then the rate slowly increases until the end of the simulation reaching values not larger than $\approx 3\times 10^{-10}\,M_{\odot}$~yr$^{-1}$. In the corresponding simulation with flares (run FL-HD) the accretion starts much earlier, about 20 hours after the start of the simulation, with rates more than an order of magnitude larger than in run NF on the timescale covered. The disk perturbation and the mass accretion observed in simulations with flares therefore are due to the flaring activity in proximity of the disk.

We compared the accretion rates derived from our 3D simulations with those available in the literature derived from optical–UV observations. In particular, we considered two samples of low-mass young accreting stars analyzed by \cite{2008ApJ...681..594H} and observed with the Low Resolution Imaging Spectrometer (LRIS) on Keck I and with the Space Telescope Imaging Spectrograph (STIS) on board the Hubble Space Telescope, and an X-ray selected sample of CTTSs analyzed by \cite{2011A&A...526A.104C} and observed with various optical telescopes. We found that the three simulations considered reproduce quite well most of the accretion rates measured in low-mass stars and they are lower than those of fast-accreting objects such as BP Tau, RU Lup or T Tau. On the other hand, it is worth noting that the timescale explored here is rather short ($\approx 80$~hours) so that, for instance, the disk viscosity has a negligible effect on the dynamics.

\subsection{Hotspots}
\label{spots}

The disk material flowing through the accretion columns is expected to impact onto the stellar surface and to produce there shock-heated spots. We do not have enough spatial resolution in our simulations to model these impacts (see, for instance, \citealt{2010A&A...510A..71O, 2013A&A...559A.127O}). Also, the model does not include a description
of the chromosphere of the CTTS, necessary to produce shocks at the
base of accretion columns after the impacts. Nevertheless, we can
infer from the simulations the expected size and evolution of the
hotspots, and the latitudes where the spots are located.

\begin{figure}
        \centering
      \includegraphics[width=9cm]{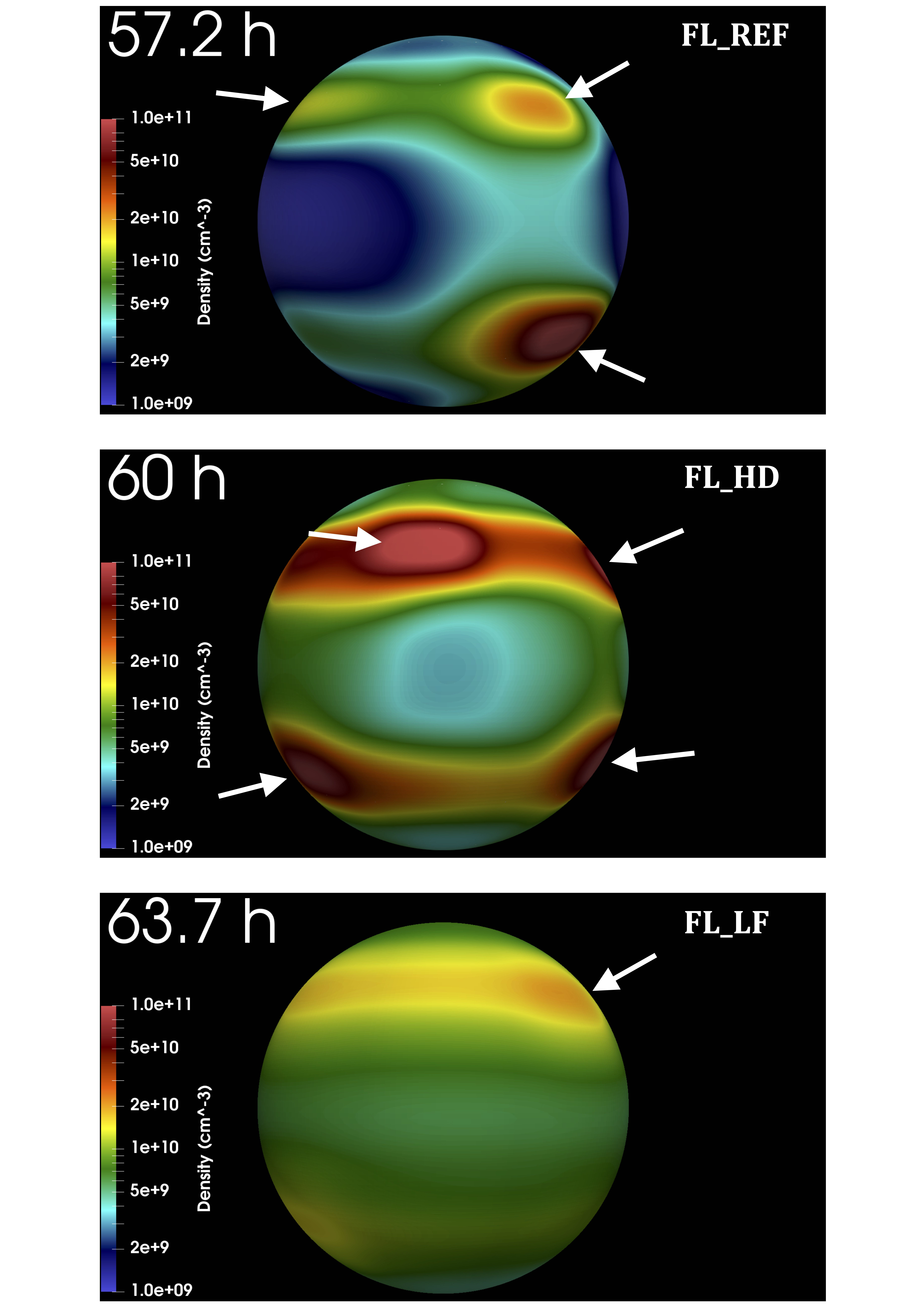}
 \caption{Maps of particle number density, in log scale, close to the
	  stellar surface after $\approx 60$~hours of evolution for
	  runs FL-REF (upper panel), FL-HD (middle panel), and FL-LP
	  (lower panel). The high density regions (pointed by white arrows) correspond to
	  accretion columns impacting onto the stellar surface.
	  Note that runs FL-REF and FL-HD exhibit clear hotspots,
	  whereas in run FL-LF the matter is more diffused and less
	  constrasted in density.}
\label{hotspot}
\end{figure}

Fig.~\ref{hotspot} shows maps of particle number density close to
the stellar surface after about 60 hours of evolution (namely when
the accretion columns are well developed). A movie showing the time evolution of particle number density  on the stellar surface is available as on-line material\footnote{The movie shows the evolution for FL-REF of the particle number density, in log scale, close to the stellar surface. } (Movie 2). During the initial
transient phase, only a few (one or two) small and faint hotspots are present on the stellar surface, each corresponding to an accretion column (Fig.~\ref{RFC} D-E and Movie 1). At later times, the number of accretion columns, in general, increases following the disk perturbation due to the flares (Fig.~\ref{RFC} F and Movie 1 ). The number of hotspots, however, does not increase with the number of accretion columns and no more than four spots are visible on the stellar surface during the whole evolution (see Fig.~\ref{hotspot}). In fact, the accretion streams interact with each other, often merging together to form larger accretion columns before impact (Fig.~\ref{RFC} F and Movie 1). This complex dynamics are typical of accretion in the strongly unstable regime, where smaller tongues merge together and form one or two larger tongues (e.g. \citealt{2016MNRAS.459.2354B}). As a result, the spots have larger size at later times.

In the reference case (FL-REF), two prominent hotspots are evident
in the visible hemisphere of the star (top panel of Fig.~\ref{hotspot})
both with density around $3 \times 10^{10}$ cm$^{-3}$ (the shock-heated plasma after impact is expected to be four times denser in the strong shock conditions; e.g. \citealt{2010A&A...522A..55S}).
A slightly fainter (density around
$10^{10}$ cm$^{-3}$) spot is also present in the northern hemisphere.
The spots cover a significant fraction of the stellar surface ($\approx 20$\,\%). Simulation FL-HD (middle panel
of Fig.~\ref{hotspot}) shows a similar density structure on the
stellar surface. In this case three large hotspots with density
ranging between $10^{10}$ and $5\times10^{10}$ cm$^{-3}$ are present
in the visible hemisphere of the star. In this case, the filling
factor of the spots is $\approx 30$\,\%. Run FL-LF presents a density
structure on the stellar surface which is significantly different
than in the other two cases. In fact, the density contrast of the
spots is much smaller than in runs FL-REF and FL-HD. A main hotspot
with density $\rho \approx 2 \times 10^{10}$ cm$^{-3}$ and two less
dense ($\rho \approx 10^{10}$ cm$^{-3}$) and smaller spots are present
in the visible hemisphere. In this case, the filling factor of the
spots is less than $10$\,\%.

Fig.~\ref{hotspot} also shows that the hotspots are characterized
by a dense inner core surrounded by lower density material as
found by \cite{2004ApJ...610..920R} (see also
\citealt{2014ApJ...795L..34B}). The spots are clearly visible in
runs FL-REF and FL-HD (yielding, in general, the higher accretion
rates); in run FL-LF, the accreted matter is more diffused in the
whole stellar surface and the spots are less contrasted in density.
In all the explored cases, the accretion occurs preferentially at high latitudes and the hotspots are located above (below) $\ang{40}$ ($-\ang{40}$); this is an effect of the magnetic field configuration chosen for the model (see Sect. \ref{model} and, e.g., \citealt{2011MNRAS.411..915R}). 
More specifically, high latitudes hotspots are favored when dipole and octupole components are parallel (e.g. \citealt{2011MNRAS.411..915R}); conversely, if dipole and octupole components are antiparallel low-latitude spots are favored (e.g. \citealt{2011MNRAS.417.1747D}). Earlier 3D simulations of accretion onto stars with dipole-plus-octupole aligned components (as in our case) have shown that the octupole component dominates in driving matter closer to the star, guiding the flows towards the high-latitude poles, and part of the matter into octupolar belts/rings (e.g. \citealt{2008MNRAS.386.1274L, 2011MNRAS.413.1061L, 2011MNRAS.416..416R, 2011MNRAS.411..915R}). 
As a result, the shape and position of the hot spots differ from those expected in a pure dipolar case, where the spots are typically observed at intermediate latitudes. In fact, in our simulations, we observe that, initially, the matter flows into accretion streams guided by the dipole magnetic field. Then, the trajectory of the flows is sligthly redirected in proximity of the star, at distances below $\approx 1.7 R_\star$, by the dominant octupole component.

\subsection{Limits of the model}
Our setup is analogous to others which have been largely used in the literature  \citep[e.g.][]{2002ApJ...578..420R,2011MNRAS.415.3380O} and is appropriate to describe the stellar-disk system in the context of young accreting stars. Nevertheless, a number of simplifications and assumptions have been made that we discuss in the following.	

First, many works \citep[e.g.][]{2010RPPh...73l6901G, 2011AN....332.1027G} suggest that a stellar magnetic field can have a more complex configuration than the one adopted for this work, namely an aligned octupole-dipole. Also, in our setup, the initial magnetic field configuration does not account for the twisting and expansion of the field lines driven by the differential rotation of the disk and the different rotation period of the disk with respect to the star. A more complex magnetic field configuration is expected to change the dynamics of accretion streams described here, leading to a possibly more complex pattern of accretion. Moreover, we expect that the flares are generated by phenomena like magnetic reconnection (see Sect. 1). For this reason, in the region where the magnetic field may generate a heat release, we expect an even more complex configuration than that assumed in our simulations. Then, the magnetic confinement of the hot plasma might be more efficient than in our simulations; this should enhance the effects of flare energy deposition on the system.

The description of the viscosity of the disk is simplified. Simulations of MRI suggests that the disk viscosity can be highly inhomogeneous and depending in time \citep{2003ARA&A..41..555B}.
However the timescale considered here (80 hours) is negligible compared to the viscosity timescale \citep[several stellar periods, e.g.][]{ ,2002ApJ...578..420R, 2009A&A...508.1117Z}, so, we do not expect to observe any effects considering a more realistic viscosity. 
Lastly, the description in terms of geometry and duration of the heat release that generate the single flare is idealized (see Sect. \ref{Qterm}). 
We expect that the geometry of the heat release strongly depends on the specific magnetic field configuration in the site where the energy release occurs.  
Moreover, in realistic conditions we do not expect that each heat release would have the same duration as we have assumed. 
Nevertheless, here we are not interested in a detailed study of the flare evolution, but only in its consequences, in terms of dynamical perturbation of the disk. From this point of view the flares simulated produce realistic perturbations.

\section{Summary and Conclusions}

We investigated the effects of an intense flaring activity localized in proximity of the accretion disk of a CTTS. To this end, we adopted the 3D MHD model, presented in Paper I, a magnetized protostar surrounded by a Keplerian accretion disk. The model has been adapted to include a storm of flares with small-to-medium intensity occurring in proximity of the disk. The model takes into account all the relevant physical processes: the stellar gravity, the viscosity of the disk, the thermal conduction, the radiative losses from optically thin plasma, and a parameterized heating function to trigger the flares.
We explored cases with different density of the disk and different levels of flaring activity. Our results lead to the following conclusions.

\begin{itemize}
\item The coronal activity due to a series of small-to-medium flares
occurring in proximity of the disk surface heats up the disk material
to temperatures of several million degrees. Part of this hot plasma
is channelled and flows in magnetic loops which link the inner part of the
disk to the central protostar; the remaining part of the heated plasma
is poorly confined by the magnetic field (especially in flares
occurring at larger distances from the star) and escapes from
the system, carrying away mass and angular momentum. The coronal
loops generated by the flares have typical lengths of the order of $10^{11}$~cm, maximum
temperatures ranging between $10^8$ K and $10^9$ K, and a lifetime
of about 10~hours. These characteristics are similar to those
derived from the analysis of luminous X-ray flares observed in young
low-mass stars (e.g. \citealt{2005ApJS..160..469F}). The escaping
disk material contributes to increase the plasma outflow from the
system with mass loss rates $M_{loss} = 10^{-10}$, and angular momentum loss rate $L_{loss} = 10^{6}-10^{7}$g cm$^{-2}M_{\odot}^{-1}$~yr$^{-1}$. The disk material, either
confined in loops or escaping from the star, forms a tenuous hot
corona extending from the central protostar to the inner portion
of the disk.

\item The circumstellar disk is heavily perturbed by the flaring
activity. In the aftermath of the flares, disk material evaporates
in the outer stellar atmosphere under the effect of the thermal conduction. 
As previously stated in Paper I, overpressure waves are generated, by the heat pulses, in the disk at the footpoints of the hot loops forming the corona. The overpressure waves travel through the disk distorting its structure. Eventually the overpressure waves can reach the side of the disk opposite to where heat pulses were injected.
There, the overpressure waves can push the disk material out of equilibrium to form funnel flows which accrete disk material onto the protostar. 
We found that the effects of the overpressure waves are larger in disks less dense and for higher frequency of flares. 
This kind of accretion process starts about 20 hours after the first heat pulse, namely a timescale much shorter than that required by the disk viscosity to trigger the accretion. The accretion rates derived by the simulations range between $10^{-10}$ and $10^{-9}\,M_{\odot}$~yr$^{-1}$. We found that the higher the disk density, the higher the accretion rate. The accretion rates are comparable with those inferred from observations of low-mass stars and brown dwarfs and in some solar mass accretors \citep{2008ApJ...681..594H,2011A&A...526A.104C}.
\item The accretion columns generated by the flaring activity on the disk have a complex dynamics and a lifetime ranging between few hours and tens of hours. They can be perturbed by the flaring activity itself; for instance a flare occuring close to an accretion column can disrupt it or, otherwise, increase the amount of downfalling plasma. The streams can also interact with each other if they are sufficiently close, possibly merging together in larger streams. As a result of this complex dynamics, the streams are highly inhomogeneous, with a complex density structure, and clumped, with clump of typical size comparable to the section of the accretion column. In general, they have a dense inner core surrounded by lower density material as also found by \cite{2004ApJ...610..920R}. These findings support the idea that persistent low-level hour-timescale variability observed in CTTSs reflect the internal clumpiness of the streams (\citealt{1996A&A...307..791G, 1998ApJ...494..336S, 2007A&A...463.1017B, 2007A&A...475..891G, 2009ApJ...706..824C}).

\end{itemize}

%

Our simulations open a number of interesting issues. The hot plasma generated by the flares produces a hot extended corona which irradiates strongly X-rays with a different spatial distribution with respect to the standard scenario of a corona confined in the stellar surface. A  corona above the disk irradiates the disk from above with near
normal incidence; the radiation reaches outer regions of the disk
that may be shielded from the stellar-emitted X-rays. This may influence in a different way the chemical and physical evolution of the disk, with important consequences, for instance, on the formation of planets. This effect can be even more relevant than that due to X-ray emission from the base of protostellar jets located at larger distances from the disks \citep[hundreds of astronomical units;][]{2007A&A...462..645B}. In addition, the irradiation produced by the flares can also increase the level of ionization of the disk. As a result, a better coupling between magnetic field and plasma can be realized, thus possibly increasing the efficiency of MRI. 

\section*{Aknowledgement}
We thank the referee for their comments that helped us to improve the manuscript. PLUTO is developed at the Turin Astronomical Observatory in
collaboration with the Department of Physics of Turin University.
We acknowledge the "Accordo Quadro INAF-CINECA (2017)” the CINECA
Award HP10B1GLGV and the HPC facility (SCAN) of the INAF – Osservatorio
Astronomico di Palermo, for the availability of high performance
computing resources and support.  This work was supported by the
Programme National de Physique Stellaire (PNPS) of CNRS/INSU co-funded
by CEA and CNES. This work has been done within the LABEX Plas@par
project, and received financial state aid managed by the Agence
Nationale de la Recherche (ANR), as part of the programme
"Investissements d'avenir" under the reference ANR-11-IDEX-0004-02.

\bibliographystyle{aa}
\bibliography{bib}

\end{document}